# Magnetic fields produced by electric railways

Martín Monteiro(a), Giovanni Organtini(b) and Arturo C. Marti(c),
(a) Universidad ORT Uruguay; monteiro@ort.edu.uy
(b) Universida di Roma La Sapienza, Italy, Giovanni.Organtini@roma1.infn.it
(c) Universidad de la República, Uruguay, marti@fisica.edu.uy

We propose a simple experiment to explore magnetic fields created by electric railways and compare them with a simple model and parameters estimated using easily available information. A pedestrian walking on an overpass above train tracks registers the components of the magnetic field with the built-in magnetometer of a smartphone. The experimental results are successfully compared with a model of the magnetic field of the transmission lines and the local Earth's magnetic field. This experiment, suitable for a field trip, involves several abilities, such as modeling the magnetic field of power lines, looking up reliable information and estimating non-easily accessible quantities.

**Electromagnetic fields.** We live surrounded by electromagnetic fields covering all ranges of spatial and temporal scales. Most of them are difficult to measure or even to detect. Notable exceptions are static or quasi-static magnetic fields which can be detected or measured with a compass or with a smartphone sensor. In the past, smartphone experiments were proposed using small currents [1-3], magnets [4,5], and the Earth magnetic field [6]. The other example, not so obvious, consists in the magnetic fields produced by power electric currents. Most of the power systems are based on alternating currents (AC), 50 Hz or 60 Hz and, as a consequence, the created magnetic field results difficult to measure. However, some railways electrification systems are powered by direct currents (DC) and thus the magnetic field produced by overhead lines (also known as catenaries) can be occasionally measured. In this work, we propose the analysis of the magnetic field produced by electric railways near Rome.

**The experiment.** Modern smartphones usually possess built-in magnetometers that have been employed in several physics experiments [1-6]. The experiment proposed here was performed outdoors in a peaceful place near Rome. A pedestrian walking on an overpass above the tracks of the railways (shown in Fig.1) registers the magnetic field with her/his smartphone. At that moment, no train was in the vicinity and no other vehicle was passing by. While the experimenter was walking at nearly constant speed (~ 1 m/s) the smartphone was held horizontally with the screen oriented upwards. This procedure is similar to the "fly-by" on an air track proposed in [3]. Thanks to the Phyphox app [7] the three components of the magnetic field were registered as the pedestrian walked over the bridge.

**The comparison.** The magnetic field is the sum of the contributions of the magnetic fields of the right track (RT), left track (LT), catenary (C) and Earth (E), $\vec{B} = \vec{B}_{RT} + \vec{B}_{LT} + \vec{B}_C + \vec{B}_E$ , with the currents and distances sketched in Fig. 2. According to the train company the power of the engines is 2.2 MW and the voltage of the lines is 3kV (DC), resulting an intensity of approximately I = 733 A, though this value can vary depending on the number of convoys nearby and their accelerations. We also assume a current in one direction through the catenary and the opposite, returning, current uniformly distributed in the two tracks. The orientation of the tracks, the path and the Earth's magnetic field were obtained from the satellite view and the reference frame is chosen to be the same of the smartphone, with the x-axis parallel to the tracks, the y-axis perpendicular to them and parallel to the overpass and the z-axis vertical.

The contributions produced by the currents are given by the Biot-Savart law for an infinite line

$$\vec{B} = \frac{\mu_0 i(-z)}{2\pi r^2}\hat{y} + \frac{\mu_0 i(y)}{2\pi r^2}\hat{z}$$

while from NOAA's geomagnetic calculator for that location, the components of Earth's magnetic field were obtained (see Fig. 1). Then, the components along the axis shown in Figs. 1-2 can be written as

$$B_y = \frac{\mu_0}{2\pi}\left(\frac{I}{2}\right)\frac{(-z)}{(y-a)^2+z^2} + \frac{\mu_0}{2\pi}\left(\frac{I}{2}\right)\frac{(-z)}{(y+a)^2+z^2} + \frac{\mu_0}{2\pi}(-I)\frac{(h-z)}{y^2+(z-h)^2} + B_{EH}\cos(\theta-\delta)$$

and

$$B_z = \frac{\mu_0}{2\pi}\left(\frac{I}{2}\right)\frac{(y-a)}{(y-a)^2+z^2} + \frac{\mu_0}{2\pi}\left(\frac{I}{2}\right)\frac{(y+a)}{(y+a)^2+z^2} + \frac{\mu_0}{2\pi}(-I)\frac{(y)}{y^2+(z-h)^2} + B_{EZ}$$

where a is the separation between tracks, h the height of the catenary and z the height of the smartphone whose values are indicated in the figure captions.

The left panel of Fig. 3 shows the experimental results obtained by the walking experimenter while the right panel plots the magnetic field corresponding to the model described above. The agreement between the field measurements and the model is clearly manifested.

**Systematic effects.** Despite smartphone's magnetometers are not so accurate [8], the measurement can be quite precise and need some precautions. The current flowing on a train line is not constant. The measurement, then, must not last too much to avoid spotting sudden changes in the current. Moreover, the measurement must be done far from large ferromagnetic volumes, such as the wagons of the train or cars passing nearby.

**Conclusion.** The magnetic field produced by the railways provides the possibility to explore the electromagnetic fields that surrounds us. Measurements can be successfully compared with estimations based on reasonable data and information available on the internet. This experiment encourage student to go outdoors and experiment using everyday tools.

**FIGURES**

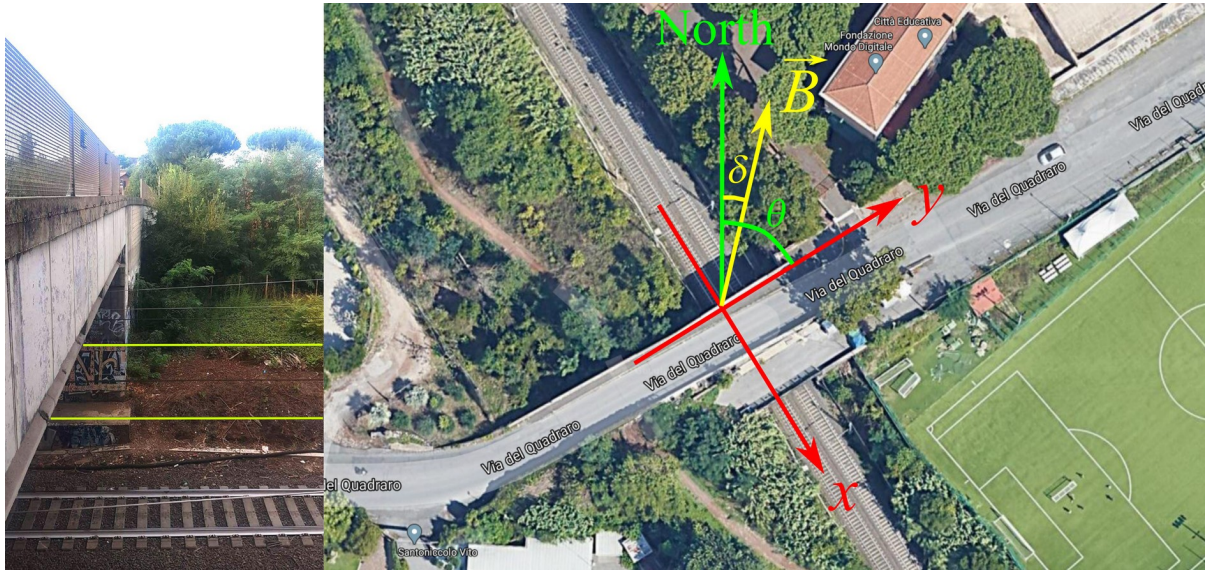

Figure 1. The magnetic field measurement was taken by a pedestrian walking on an overpass above the tracks (left panel) shown in the satellite view (right panel). Axis $x$ points in the direction of the currents through the tracks, opposite to the current $I$ through the catenary (highlighted in yellow) while $y$ is along the smartphone's path, and $z$ points vertically upwards. From NOAA website we obtained the horizontal component of Earth's magnetic field, $B_{EZ}$ = - 24.6 µT, the magnetic declination, $\delta$ = 3.45º and the vertical component of Earth's magnetic field $B_{EH}$= -39.6 µT. From map, angle $\theta$ = 61.9º.

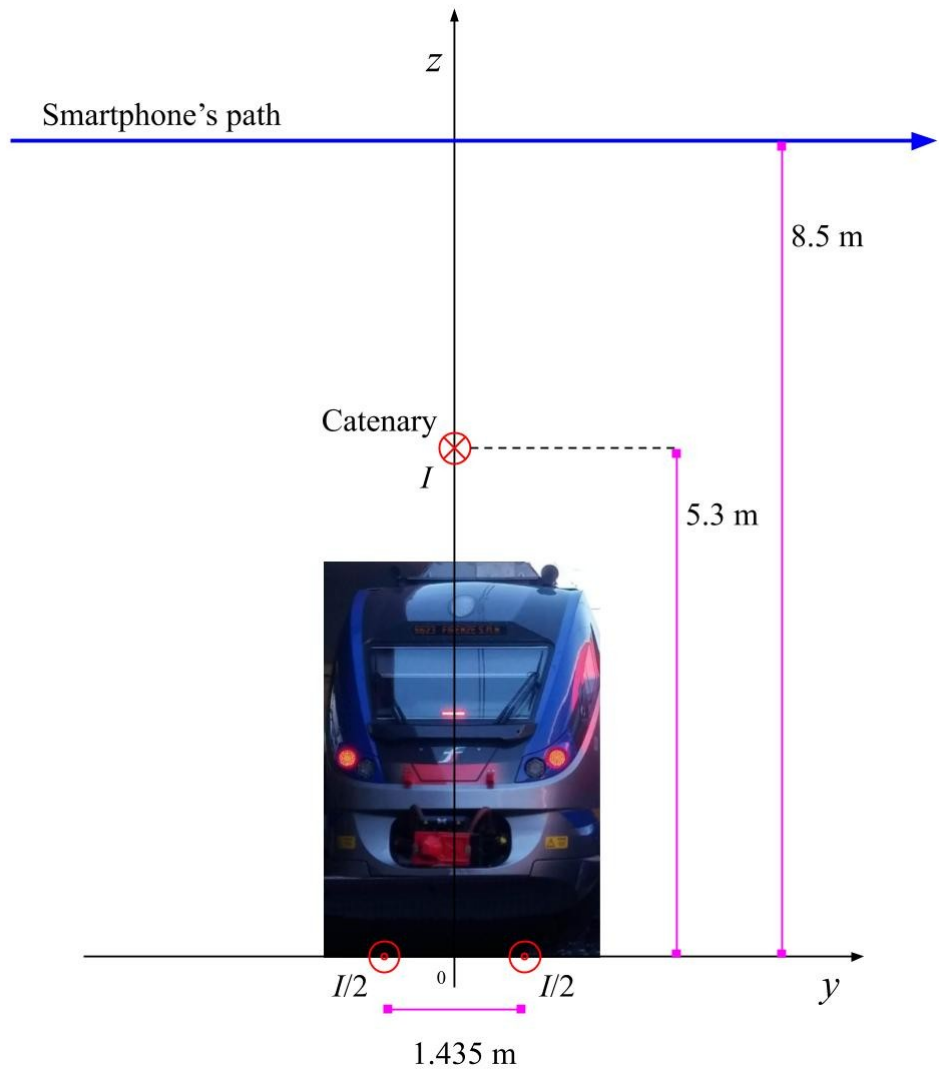

**Figure 2.** Layout of the currents and the smartphone's path. The standard gauge railways and height of the catenary were taken from the website of the company while the height of the bridge was estimated from the pictures.

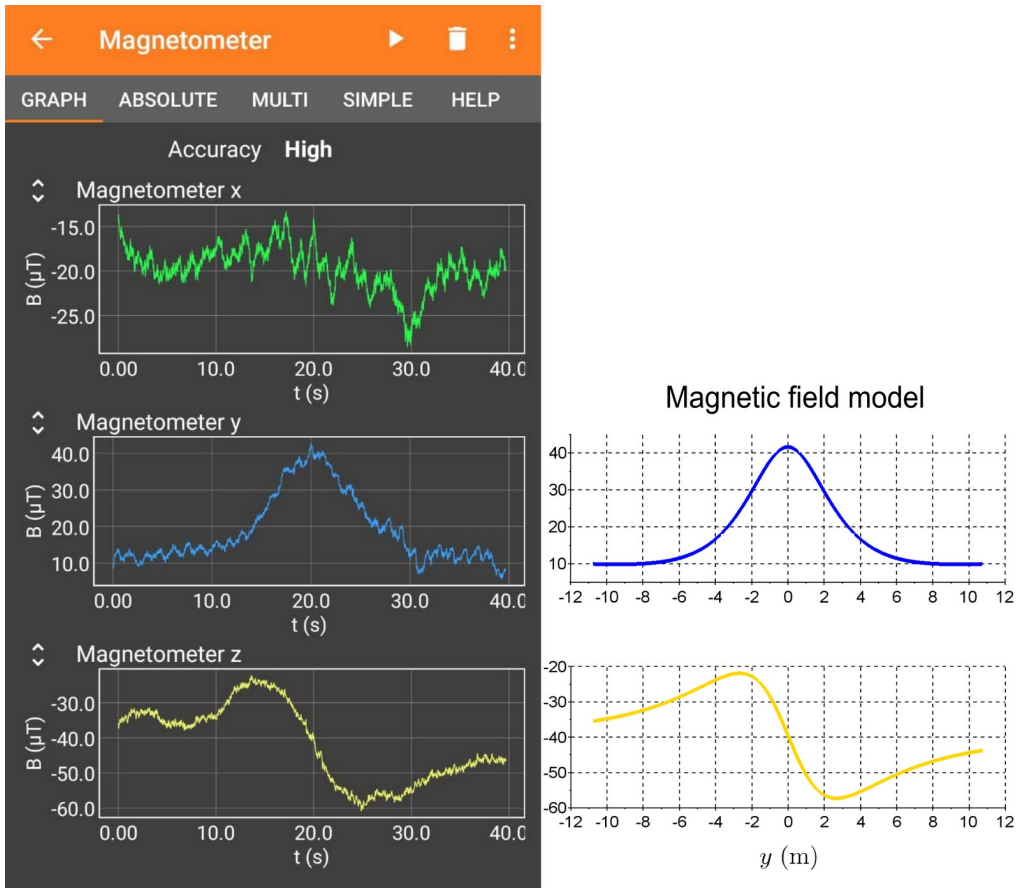

**Figure 3.** Comparison between the measurements (Phyphox screenshot on the left) and the model calculation (on the right).